\documentclass[a4paper,floatfix,preprint,superscriptaddress]{revtex4}
  \usepackage[latin1]{inputenc}
  \usepackage{graphicx}
  \usepackage{booktabs}

  \usepackage{latexsym}
  \usepackage{amssymb}
  \usepackage{amsmath}
  \usepackage{amsfonts}
  \usepackage[english]{babel}
  \usepackage{ccaption}

\begin{document}
  \providecommand{\MIT}{Department of Physics, Massachusetts Institute of Technology, USA. }
  \providecommand{\ICREA}{ICREA-Complex Systems Lab, Universitat Pompeu Fabra, 08003 Barcelona, Spain. }
  \providecommand{\IBE}{Institut de Biologia Evolutiva (CSIC-UPF), 08003 Barcelona, Spain. }
  \providecommand{\USC}{Departamento de F\'isica Aplicada, Universidade de Santiago de Compostela, 15782
Santiago de Compostela, Spain. }

  \author{Lu\'is F Seoane} \affiliation{\MIT} \affiliation{\ICREA} \affiliation{\IBE}
  \author{Jorge Mira} \affiliation{\USC}

  \title{Modeling the life and death of competing languages from a physical and mathematical
  perspective\footnote{Reference: Seoane LF and Mira J. Modeling the life and death of competing languages from a physical and mathematical perspective. In {\em Bilingualism and Minority Languages in Europe: Current trends and developments} (F. Lauchlan, M. C. Parafita Couto, eds.), pp.70-93 (ISBN: 978-1-4438-1943-5). Cambridge Scholars Press (2017). }}

  \date{\today}

  \begin{abstract}
 
    Recent contributions address the problem of language coexistence as that of two species competing to aggregate speakers, thus focusing on the dynamics of linguistic traits across populations. They draw inspiration from physics and biology and share some underlying ideas -- e. g. the search for minimal schemes to explain complex situations or the notion that languages are extant entities in a societal context and, accordingly, that objective, mathematical laws emerge driving the aforementioned dynamics. Different proposals pay attention to distinct aspects of such systems: Some of them emphasize the distribution of the population in geographical space, others research exhaustively the role of bilinguals in idealized situations (e. g. isolated populations), and yet others rely extremely on equations taken unchanged from physics or biology and whose parameters bear actual geometrical meaning. Despite the sources of these models -- so unrelated to linguistics -- sound results begin to surface that establish conditions and make testable predictions regarding language survival within populations of speakers, with a decisive role reserved to bilingualism. Here we review the most recent works and their interesting outcomes stressing their physical theoretical basis, and discuss the relevance and meaning of the abstract mathematical findings for real-life situations.

  \end{abstract}

  \maketitle

  \section{Introduction}
    \label{sec:1}

    Language is a defining trait and a prominent way of communication for the human species \cite{Bickerton1992, MaynardSmithSzathmary1997}. On the other hand, languages are social constructs with an entity of their own \cite{DyenBlack1992, PetroniServa2008}. These emergent objects, the shared codes that allow communication, are our concern throughout this paper. We talk about contact situations when speakers of different tongues come together. They decide what language to adopt when talking to each other, and this leads to certain dynamics of use and knowledge across populations. These dynamics reflect an {\em interaction} between languages. Diverse phenomenology exists associated to such contact situations \cite{McMahonMcMahon2005}, but we are interested in the coarse grained dynamics of the number of speakers of different (usually two) languages within a population. Contact situations happen for social constructs other than languages \cite{CastellanoLoreto2009}. Opinions and political views \cite{SanKlemm2005, FernandezEguiluz2014} are shared and discussed while economic agents engage in fusions, competitions, etc, that make them rise and fall \cite{DaeppBettencourt2015}. These interactions are resolved by the choices made by individuals, who are free to take sides or change their minds over time. This determines the fate of the emergent social objects preventing or giving rise to homogeneous opinions, political shifts, strong economic holdings, or unified communities of speakers.

    A legitimate question is whether there exist mathematical rules behind these interactions and whether some universal regularities affect every human society alike. A premise behind this question is that, while we cannot track each individual, aggregating large masses might dissolve the most skewed positions revealing clear-cut rules for the average trends \cite{Ball2002, Ball2004, Ball2012}. This picture is fully exploited by the branch of physics known as statistical mechanics. To solve the dynamics going on in a glass of water we need to keep track of the location, velocity, and energy of every water molecule. These numbers vary greatly as atoms bump into each other and exchange energy and momentum -- i.e. as they interact -- so the task readily becomes impossible. But statistics ensures that if we average millions of molecules together those large fluctuations cancel out while the averages behave according to smooth mathematical rules. Thus, the mean position, velocity, and energy become volume, pressure, and temperature; and the laws of thermodynamics tell us how these quantities relate to each other. Similarly, consider humans as a sort of {\em social atom} \cite{Buchanan2008} endowed with free will to fluctuate around some given options, or to interact with each other interchanging opinions, trading, through coercion, etc. Looking at small groups we often observe an untameable diversity. As we amass large numbers of people, trends emerge and the rate of change between options may become more regular. Mathematical laws might arise, not because humans follow some unfathomable law, but as a statistical property out of their free choices. Maximally random particles yield some of the better established physical laws -- those of thermodynamics. Given an ensemble of not-so-random agents, why would they not similarly obey some elegant mathematical laws? Interestingly, it was the development of these ideas in social sciences what brought statistics to physics \cite{Ball2002, Ball2004, Ball2012}. We are, somehow, returning to the origins.

\begin{figure}[htbp]
      \begin{center}

        \includegraphics[width = 15 cm]{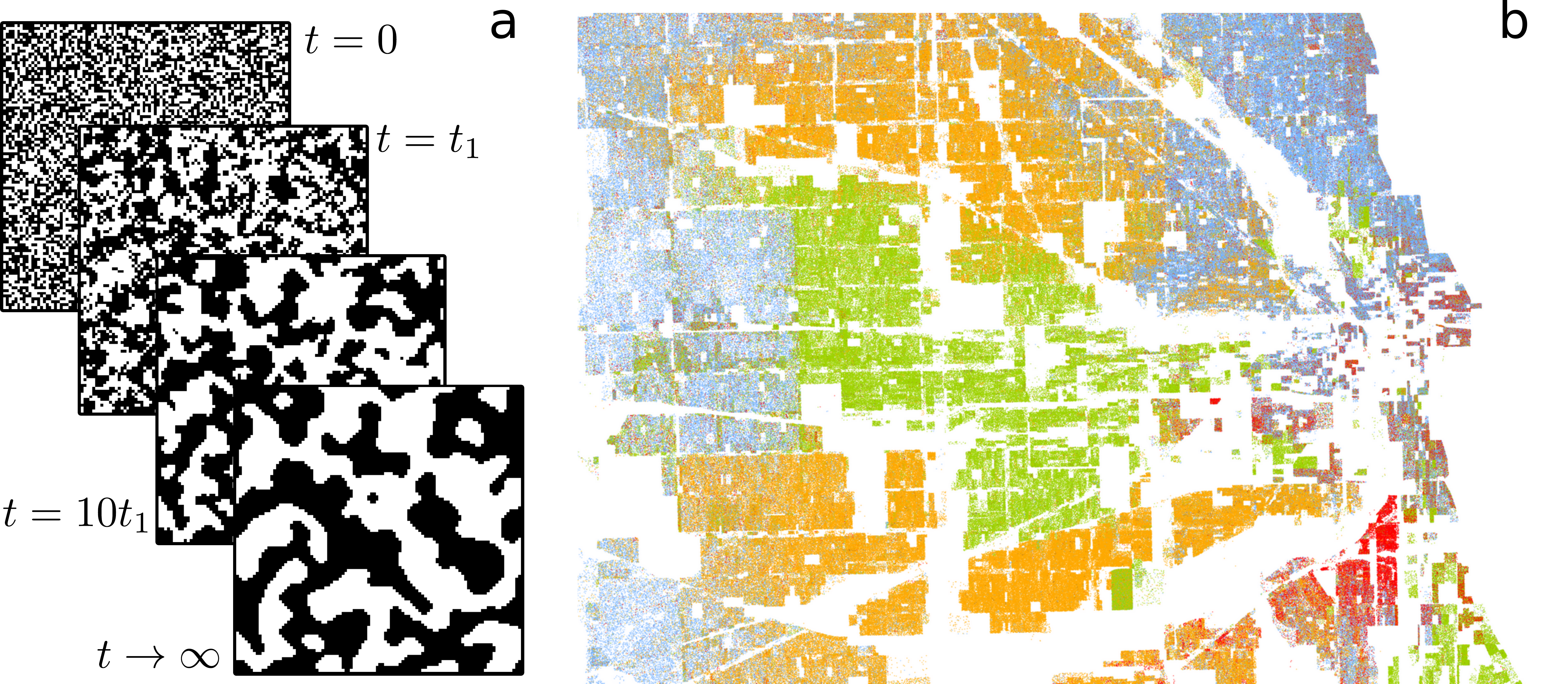}

        \caption{{\bf A model for racial segregation}. {\bf a} Four snapshots of Schelling's model, that describes ghetto formation. $t_1$ is chosen such that, in average, within $t$ and $t_1$, $1\%$ of the population has attempted to move into a new place. {\bf b} Racial segregation in Chicago. Each dot corresponds to one person. Colors indicate the ethnics: blue, White; green, Black; red, Asian; orange, Hispanic; brown others. ({\em Image Copyright, 2013, Weldon Cooper Center for Public Service, Rector and Visitors of the University of Virginia (Dustin A. Cable, creator)} -- see \cite{Cable2013}.) }

       \label{fig:1}
      \end{center}
    \end{figure}

    Early works on this field include studies of ethnic and cultural dynamics. Schelling (\cite{Schelling1971}; see also \cite{Clark1991} and \cite{Krugman1996}) proposed a simple model of citizens choosing their neighborhood. Consider a checkerboard with its squares distributed randomly (figure \ref{fig:1}{\bf a}). Each square represents a house whose occupant opts for one of two options (black or white). Every day there is a probability that each individual tries to move out into a random location. She does so if at the new location she has more similar neighbors than before. Schelling showed that, no matter how low the probability of moving out is, the checkerboard tends to maximally segregated communities. We rarely observe fully isolated races or social classes (which correspond to the black and white options of the model), but actual scenarios are close enough (figure \ref{fig:1}{\bf b}, \cite{Cable2013}). In another celebrated work, the political scientist Robert Axelrod proposed a model of how cultures survive in an interconnected world \cite{Axelrod1997}. According to this model we tend to a unified dominant culture if the level of human interaction is above a certain threshold. Disperse and segregated cultures survive if that interaction is kept low -- i.e. if cultures are effectively isolated.

    Despite these interesting results, it must always be noted the difference between the real world and what is possible within a model. Along this paper we will deal with {\em idealized} languages whose properties may often contradict the intuition of a linguist. We never claim that the results outlined apply to all languages, or that we are delivering unchangeable truths -- this is against scientific thought itself. Rather, we will try to convey in written words what the solutions to a series of equations mean for these idealized languages. This suggests constraints to the real world, given our mathematical knowledge and, sometimes, empirical data. All this must be challenged as new evidence mounts up. Whenever some notion contradicts well-established knowledge in linguistics, this is suggesting us how our models shall be improved in the future, thus allowing a greater prediction power. Take as examples the idealized situations described by Schelling and Axelrod. They establish limiting cases that we expect to find in some social systems {\em unless something else is going on}. These limits send clear messages within a model of idealized cultures: keep segregated or risk absorption by a uniforming option. Because the real world is often far from such homogeneous scenarios, we can be quite sure that {\em something else is going on} that has not been taken into account by the equations, and this suggests how to improve them.

    Language contact situations have theoretical relevance to understand universal laws of social interaction. At the same time these theoretical advances have implications for {\em minority languages} and make suggestions to help promote them. As minority languages we understand those with low prestige whose speakers are not necessarily a minority when the contact with a {\em dominating language} (larger prestige) begins. We review recent advances from the perspective of physics, which has given some theoretical approaches to language contact situations. We are concerned with the survival of minority languages, so we focus on works that have addressed specific aspects that seem critical for this issue. Works developing more technical and theoretical details are discussed in the abundant, up-to-date reviews about the topic \cite{Wichmann2008, KandlerSteele2008, Kandler2009, SoleFortuny2010, PatriarcaSan2012, CastelloSan2013}. 

    Two relevant factors have been identified that affect the survival of minority languages: geography and bilingualism. In section II we review models that gradually incorporate these elements, with emphasis on the latter. We discuss the prospects of minority languages given the theoretical results. These, as for the Schilling and Axelrod models, refer to ideal mathematical situations which are difficult to bring together with real data. We attempt to do that in section III fitting the dynamics of Galician-Spanish coexistence in northwest Spain. Obvious limitations show up, also concerning the kind of data available which is often scarce, qualitative, and difficult to link to current theories \cite{Wichmann2008, Crystal2000}. This is so because we are dealing with systems with many relevant facets that affect each other in nontrivial ways, another reason why theoretical results must be taken with extreme care. These and other issues are discussed in the concluding part, section IV.

  \section{An emerging picture: the role of bilingualism in models of language population dynamics}
    \label{sec:2}

    \subsection{The Abrams-Strogatz model}
      \label{sec:2.1}

      In 2003, Abrams and Strogatz introduced a concise, yet insightful model that captured the dynamics of dying languages \cite{AbramsStrogatz2003}. Earlier approaches existed \cite{BaggsFreedman1990, BaggsFreedman1993}, but Abrams and Strogatz's contribution was simple and their equations had clear interpretations that explained a wealth of data parsimoniously. This sparked a new wave of research that expanded their methods and lasts until today.

      Assume a population with $x$ speakers of language $X$ and $y$ speakers of language $Y$. Consider this population to be constant and normalized to unity ($x+y=1$). Let, $P_{YX}$ be the probability that a speaker decides to abandon language $Y$ for language $X$. Similarly, $P_{XY}$ for $X$-speakers switching to $Y$. This can be written mathematically in the form $P_{YX} = xs_Xx^a$ and $P_{XY} = cs_Yy^a$, where $s_X$ and $s_Y$ introduce the prestiges. The parameter $a$ was termed volatility by Castell\'o et al. \cite{CastelloSan2013}, who associate it to a tendency to switch language use.

      The constant $c$ is related to a global rate at which the language shift takes place: If $c$ is larger, the flow of speakers from one language to another occurs faster; if it is smaller, the dynamics take more time to trace the same relative evolution. The approach of the literature reviewed here often focuses on the distributions of speakers when a long time has elapsed. This is not affected by $c$, so this parameter has been largely ignored. It would be very interesting, though, if some future works would relate empirically the parameter from this model to actual situations of language coexistence.

      We take the prestiges to be normalized $s_X + s_Y = 1$ (hence $s_X \equiv s$ and $s_Y = 1-s$). The symbol $s$ here stems from the original work by Abrams and Strogatz \cite{AbramsStrogatz2003}, in which the word {\em status} was used instead of prestige. We adopt the latter term in accordance to the linguistics literature. 

\begin{figure}[htbp]
        \begin{center}

          \includegraphics[width = 10 cm]{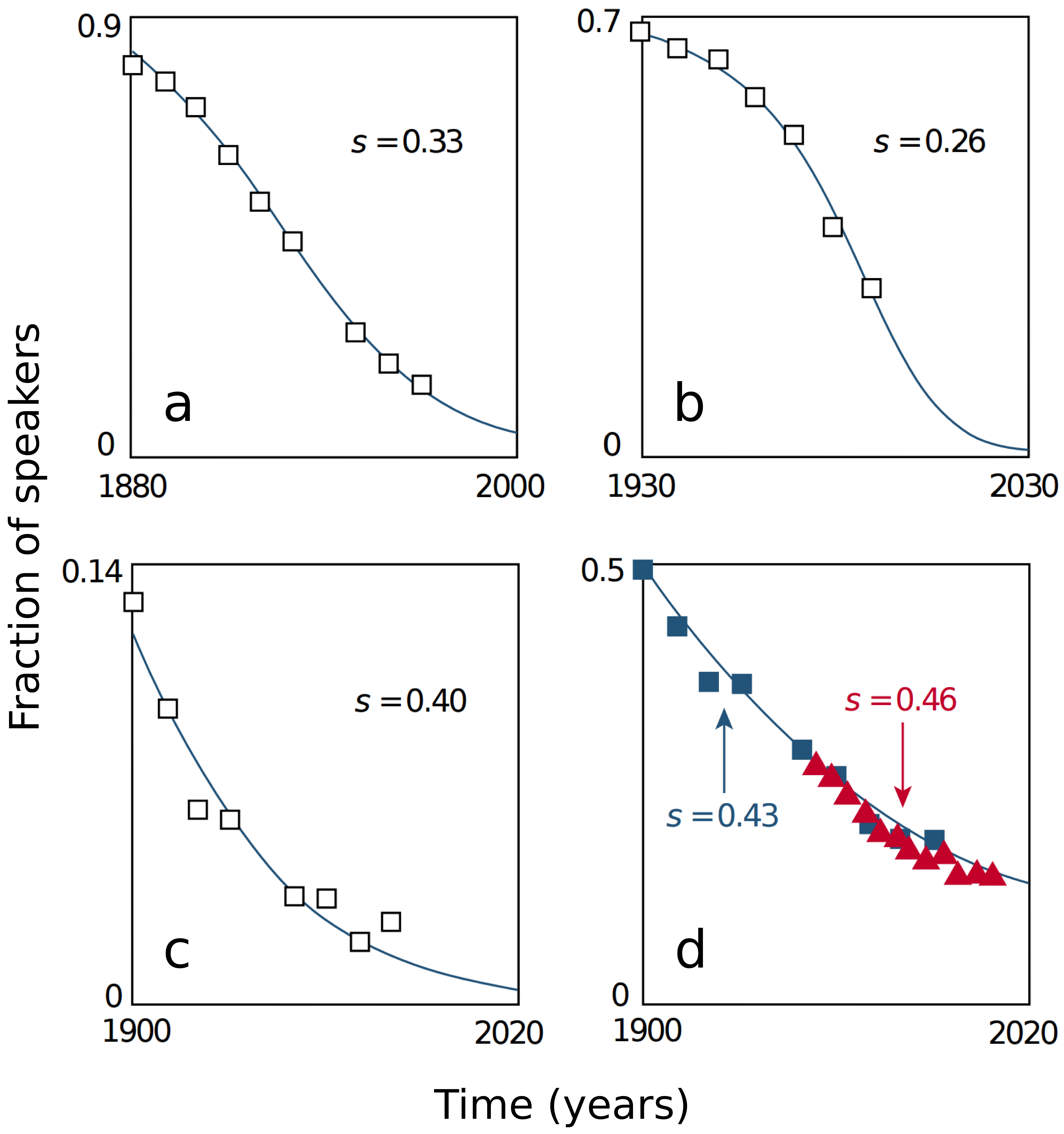}

          \caption{\textbf{Abrams-Strogatz model for language competition.} (Adapted from \cite{AbramsStrogatz2003}.) Real data from diverse pairs of competing languages ({\bf a}) Scottish Gaelic in Sutherland, Scotland; {\bf b} Quechua in Huanuco, Peru; {\bf c} Welsh in Monmouthshire, Wales; and {\bf d} Welsh in all of Wales). The solid lines indicate the predictions given by the AS equations, which, after comparison with experimental data, allow to infer the prestige parameter and to describe the decline of these languages.}

         \label{fig:2}
        \end{center}
      \end{figure}

      The probabilities $P_{XY}$ and $P_{YX}$ of switching languages build up a flux of speakers that is captured by a set of differential equations (see \cite{AbramsStrogatz2003} for details). The solutions to these equations ($x(t)$ and $y(t)$) tell us the fraction of $X$- and $Y$-speakers over time (figure \ref{fig:2}). As time tends to infinity, the system relaxes to the fixed fractions $x^*$ and $y^*$ at which the flux of speakers ceases. This is the stable state of the system. What equilibrium configuration is reached depends on the parameters ($a$ and $s$, but not $c$) and the number of speakers of each tongue when the contact process began. For this model the only stable states for $a>1$ imply the extinction of one of the languages, usually the one with lowest prestige even if it has a large initial supporting population. For $a<1$, the survival of both languages is possible \cite{AbramsStrogatz2003, VazquezSan2010}. A detailed explanation of how this volatility parameter affects the dynamics in this and other models can be found in \cite{CastelloSan2013}. 

      The power of this model is revealed when its equations are confronted with real data. To do this, recorded time evolutions of fractions of speakers are plotted against time and the equations of the model are fit to the data (figure \ref{fig:2}). As a result of this fit we obtain the most likely parameters associated to the data -- e.g. we derive the prestige parameter $s$ for each pair of language under research. Different combinations of parameters could account for the decay of up to $42$ pairs of minority languages in quite different contexts \cite{AbramsStrogatz2003}. It was consistently found that $a>1$ (indeed $a\sim1.31$, far from the $a<1$ values that allow coexistence); so that this model together with the data used to test it are bad news for minority languages: they will likely be driven to extinction unless they do not engage in competition.

      These results were rapidly challenged by more realistic equations: ``How does geography affect the competition?'' ``What if bilingualism is considered?'' But the Abrams-Strogatz (AS) model remains a remarkable contribution for two reasons: i) it primed the use of new tools in language population dynamics and ii) it established clear limiting cases similarly to how Schelling and Axelrod did for ethnic groups or cultures.

    \subsection{Geographic constraints}
      \label{sec:2.2}

      Patriarca and Lepp\"anen \cite{PatriarcaLeppanen2004} introduced the geographical factor in the AS model through a population spread over a plane with two adjacent regions. Each region ascribed a larger prestige to one of the tongues. Locally, the languages interact following the AS model. Both languages survive within their preferred region, but this segregated coexistence is due to the odd assignment of prestige. Indeed, prestige should arise from the use of language, not from properties of the land. If we substitute {\em all} speakers anywhere by speakers with a foreign language, the new people will not switch spontaneously to the old tongue. Seeking more realism, Patriarca and Heinsalu \cite{PatriarcaHeinsalu2009} study another important feature related to space: human mobility. In \cite{PatriarcaHeinsalu2009} speakers are allowed to move randomly while the prestiges of the languages do not depend on the physical location of the speakers (as they did in \cite{PatriarcaLeppanen2004}).

\begin{figure}[htbp]
        \begin{center}

          \includegraphics[width = 10 cm]{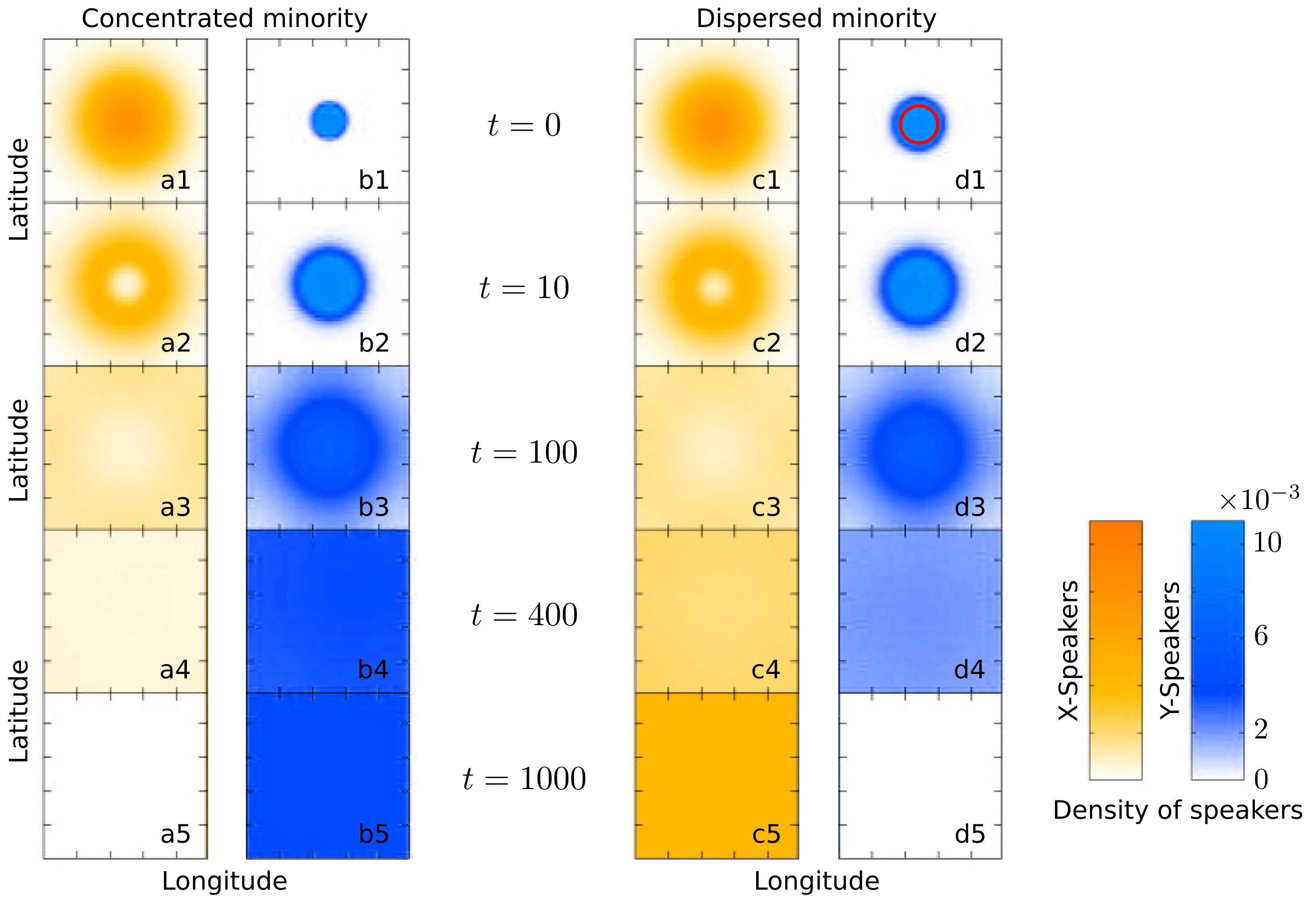}

          \caption{\textbf{Geography affects the competition.} (Adapted from \cite{PatriarcaHeinsalu2009}.) Colors represent density of speakers in a $2$-D lattice as they evolve over time. {\bf a-b} $Y$-speakers (minority language, $s_Y = 0.45$) are concentrated in a small region from where they push $X$-speakers ($s_X = 0.55$) away. Then, they successfully colonize the rest of the land driving $X$ to extinction. {\bf c-d} Same initial $X$-speakers configuration. $Y$-speakers are more diffused (red circle indicates old distribution). This is fatal for $Y$, since it cannot stand against $X$ and dies away. }

         \label{fig:3}
        \end{center}
      \end{figure}

      Figure \ref{fig:3} shows how dense communities of minority speakers might reverse an unfavorable situation \cite{PatriarcaHeinsalu2009}. $X$-speakers (dominant) are distributed equally in figures \ref{fig:3}{\bf a1} and \ref{fig:3}{\bf c1}, $Y$-speakers are more concentrated in the initial distribution of figure \ref{fig:3}{\bf b1} than in \ref{fig:3}{\bf d1}. The more compact community nucleates a resistance against $X$ that eventually turns over the situation, driving $X$ to extinction (figure \ref{fig:3}{\bf a5}, \ref{fig:3}{\bf b5}). The more diffused $Y$-speakers cannot avoid the extinction of their language (figure \ref{fig:3}{\bf c5}, \ref{fig:3}{\bf d5}).

      Survival of both languages in \cite{PatriarcaHeinsalu2009} is also possible, and it is associated to landforms such as mountains or islands. This does not depend on artificially larger prestiges (as in \cite{PatriarcaLeppanen2004}). Instead, human mobility is the key due to the low mixture imposed by the physical barriers. A clear message is sent: survival of different languages within a geographic space is possible, but both tongues must remain relatively segregated. This might be fine for the dominating language, but isolation for minority tongues seems harmful (e.g. in economic terms) or impossible, especially in our hyper-communicated world. Note that this is a result valid within the hypothesis of this model. Again, while real situations will likely be more complicated, the model establishes clear frontiers and mechanisms for the survival of coexisting tongues.

      An exploration of the model from Patriarca and Heinsalu \cite{PatriarcaHeinsalu2009} (and variations upon it) with a stress on an ecological interpretation is made in \cite{KandlerSteele2008}: $X$ and $Y$ become species competing for resources (speakers). Other models not based on the AS equations incorporate geometry as well \cite{SchulzeStauffer2007, HadzibeganovicSchulze2008} and yet other models study the diffusion of linguistic traits across space \cite{OliveiraTsang2006a, OliveiraTsang2006b, StaufferSolomon2006, SchulzeWichmann2008}. The latter focuses on the broad distributions of languages spoken in our planet, the number of tongues within a family, and the number of speakers of a tongue in average. These works tell us little about a specific contact situation, but they interpret minority languages in a systemic sense: It is possible that the distribution of languages (including endangered ones) across the planet \cite{Sutherland2003} arises as a trait of a large system in overall equilibrium.

      Finally, note that geometry is a way of constraining the interactions between speakers (modeled as abstract agents when simulating the equations computationally). In the models reviewed so far, these agents change their language depending on other agents nearby, unless some landforms prevent that. Hence, speakers interact with their neighbors in a two dimensional grid. Other kinds of constraints affect everyday real interactions -- e.g. belonging to a community such as a university or a trade union, practicing sports, etc. These interactions bias how and which agents come together, thus composing a {\em social network} that effectively alters the two dimensional geometric structure presented so far. Several works study how the underlying social network affects the dynamics of the Abrams-Strogatz models \cite{CastelloSan2013, VazquezSan2010, CastelloSan2006, CastelloSan2007a, CastelloSan2007b, StaufferSan2007}. These works also deal with bilingualism and important insights will be discussed later. Regarding monolingual models alone, these social structures seem not to alter notably the dynamics studied so far. 

    \subsection{Bilingual modifications to the Abrams-Strogatz model}
      \label{sec:2.3}

      Bilingualism is a crucial element of language contact. It could be dismissed before (as the good results in \cite{AbramsStrogatz2003} indicate) because the pairs of tongues involved are fairly different and bilingualism was relatively marginal. `Relatively marginal' does not mean that it is not important or that it does not exist. It means that it has been possible to develop a theory that satisfactorily explains some salient dynamics but that does not take bilingualism into account. When we explore language dynamics in deeper detail, evidence readily suggests that it is necessary to incorporate a notion of bilingualism into our models \cite{MiraParedes2005, KandlerSteele2010, MiraNieto2011, ZhangGong2013}. When dealing with empirical data, bilingualism is defined through some qualitative traits -- e.g. competence or usage of both languages -- as will be exemplified later. But now bilingualism is an abstract concept whose effective definition depends only on what terms it brings into the equations, and qualitative issues are less important.

      Mira and colleagues \cite{MiraParedes2005, MiraNieto2011, OteroMira2013} introduced bilingualism in the AS model through a group of bilinguals ($B$, as opposed to the previous $X$ and $Y$ monolinguals). A fraction $b$ of speakers belongs to this group. Population size remains constant and normalized ($x+y+b=1$). They considered an {\em attracting population} that includes the bilinguals, so that the urge of $Y$-speakers to learn a new language is proportional to $(x+b)^a$. The term {\em attracting population} was introduced in \cite{HeinsaluLeonard2014} and reflects what members of a society might induce a speaker to switch languages. We come back to it later. From all $Y$-speakers changing options, a fraction $k$ retain their language while $(1-k)$ become full $X$-monolinguals. This is their working definition of bilingualism: an abstract group $B$ able to attract population away from $X$ and $Y$ and capable of retaining a fraction $k$ of the shifting population. The larger $k$, the easier it is for speakers to join $B$. Hence, when the model is matched to real data and $B$-speakers correspond to a qualitative notion of bilinguals, $k$ becomes a measure of how easy or convenient it is to master both languages -- i.e. of the perceived similarity between $X$ and $Y$. This parameter was termed {\em interlinguistic similarity}.

      We can write down all probabilities of switching groups as we did for the AS model ($P_{XB} = ck(1-s)(y+b)^a$, $P_{YB} = cks(x+b)^a$, $P_{BX} = P_{YX} = c(1-k)s(x+b)^a$, $P_{BY} = P_{XY} = c(1-k)(1-s)(x+b)^a$). These again establish fluxes that are captured by a set of differential equations \cite{MiraParedes2005} . In figure \ref{fig:4}{\bf a1-a3} we plot possible dynamics that these equations can generate. What evolutions we see for a pair of languages depends on their parameters ($a$, $c$, $k$, $s$) and initial conditions. Periodic oscillations are not possible \cite{OteroMira2013} so every contact situation evolves towards stable fractions of speakers ($x^*$, $y^*$, $b^*$). Some of the possible equilibriums of this model present monolingual populations of both tongues always along a bilingual community. Hence, persistent bilingualism is possible and it enables the steady coexistence of both languages without spatial segregation \cite{MiraNieto2011}. Bilingualism acts as a contact surface between the opposed groups that softens the competition. Without bilingualism, the models most faithful to data lead either to the extinction or isolation of the minority language, as seen before. Bilingualism opens intermediate scenarios with varying (even large) presence of the minority language.

      The distribution of speakers at equilibrium ($x^*$, $y^*$, $b^*$) depends on $a$, $k$, and $s$ (but not $c$) and the initial fractions in each group when the contact situation started. In figure \ref{fig:4}{\bf b} we plot these parameters in three axes. The surfaces enclose those $(a, k, s)$ combinations for which language coexistence is possible. Figure \ref{fig:4}{\bf c} is a cross section of that surface. Empirical measurements from Galician-Spanish coexistence are plotted to assess the fate of these contact situations: according to the model, red crosses belong to systems that will lose the endangered language, while black plus symbols might present stable coexistence. Alternatively we can extrapolate the equations with the measured parameters and see if bilingualism declines or persist over time (figure \ref{fig:4}{\bf d}).

      \begin{figure}[htbp]
        \begin{center}

          \includegraphics[width = 12 cm]{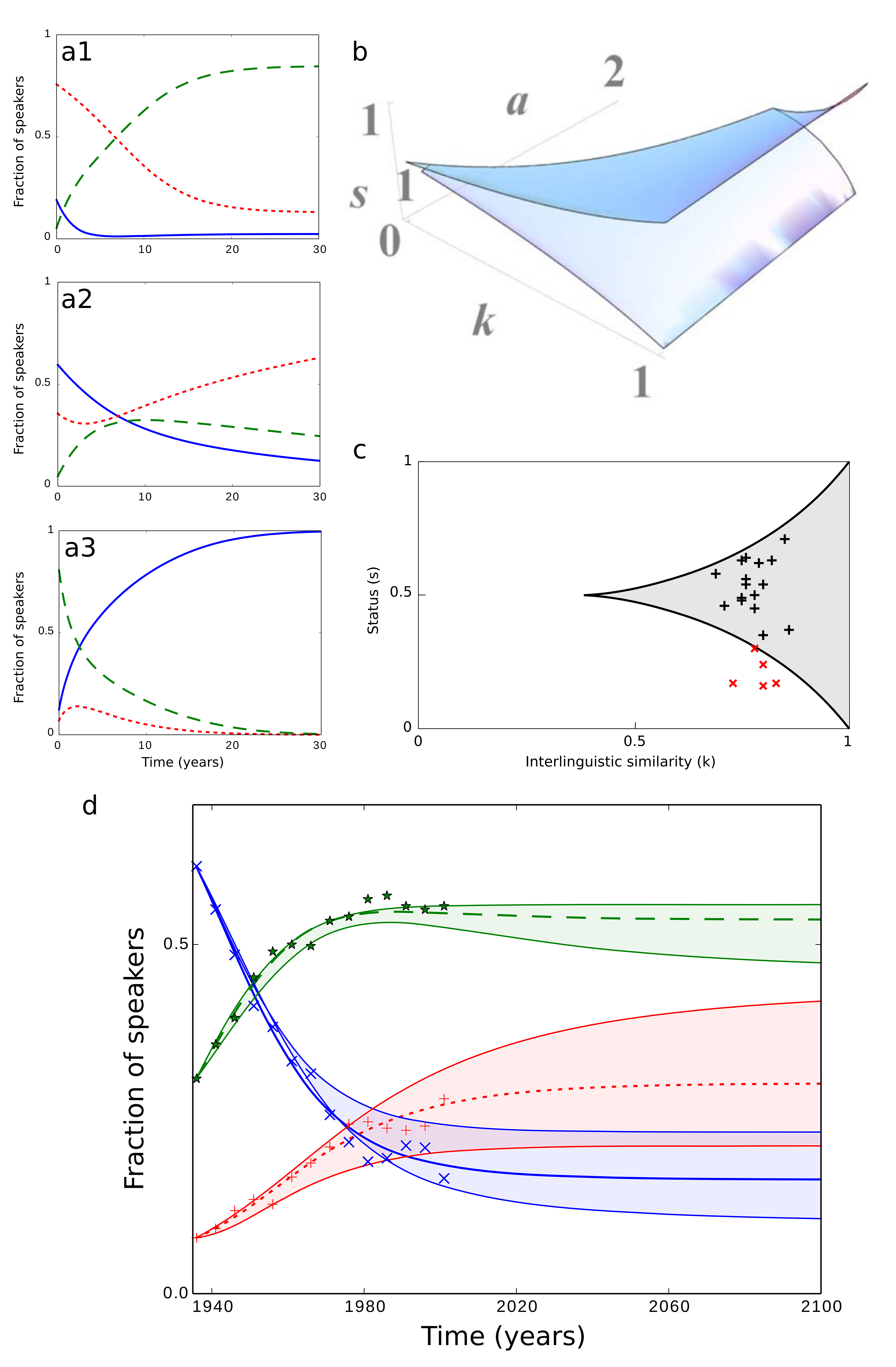}

          \caption{\textbf{A bilingual AS-model.} {\bf a} Three possible evolutions of $X$- (solid, blue), $B$- (dashed, green), and $Y$-speakers (dotted, red). Model parameters: {\bf a1} $x_0=0.2$, $y_0=0.76$, $s_X=0.32$, $k=0.94$, $a=1.31$. {\bf a2} $x_0=0.6;$, $y_0=0.36$, $s_X=0.45$, $k=0.5$, $a=1.31;$. {\bf a3} $x_0=0.1;$, $y_0=0.06$, $s_X=0.65$, $k=0.5$, $a=1.31$. (Continues in the next page.) }

         \label{fig:4}
        \end{center}
      \end{figure}
      \begin{figure}[t]
        \contcaption{(Continues from the previous page.) {\bf b} (Adapted from \cite{ColucciOtero2014}.) For parameters within the sheets, stable distributions exist that allow both languages to survive along a bilingual group. {\bf c} A cross section with $a=1.31$ allow us to compare the prestige and interlinguistic similarity of pairs of languages engaged in real competition. The gray area indicates that stable coexistence is possible, red crosses correspond to unstable language coexistence. {\bf d} The evolution of Galician (solid blue, crosses), Spanish (dotted red, plus symbols), and Bilingual (dashed green, stars) speakers in Galicia (Spain) has been reproduced by the model in \cite{MiraParedes2005} with $a=1.31$, $k=0.71$, and $s=0.46$, which is compatible with a stable coexistence. ({\bf c} and {\bf d} adapted from \cite{SeoaneMiraInPrep}). Data was obtained from \cite{RAG2004, RAG2008}. }
      \end{figure}

      Earlier attempts to model bilingualism by Baggs and Freedman \cite{BaggsFreedman1990, BaggsFreedman1993} in the 1990's had already advanced the possibility of stable language coexistence. Their models were fairly abstract and required further assumptions (as in \cite{BaggsFreedman1993}) before they could reproduce realistic scenarios. The Baggs-Freedman models broke the symmetry between $X$ and $Y$, explicitly favoring a dominating tongue. Since coexistence is still possible even so, bilingualism again means good news for minority languages.

      In a remarkable model, Zhang and Gong \cite{ZhangGong2013} rely solely on biological and physical principles to deduce equations that, with minimal empirical input, faithfully reconstruct several real situations. They incorporate bilingualism naturally. Failing to do so would not yield good reconstructions of English-Welsh and English-Gaelic competitions, the authors find, thus stressing the relevance of bilingualism. This is a promising approach that derives relevant parameters from first principles and that shall apply where little data is available. The parameters involved have clear interpretations in terms of influence between languages and population dynamics. Those equations are also ready to adopt more languages and geography appears naturally. We look forward to more applications of this model in relevant, ongoing contact situations.

      Minett and Wang \cite{MinettWang2008} also include bilingualism in the AS model. They only allow $\{X, Y\} \leftrightarrow B$ (not $X \leftrightarrow Y$) flows and take monolinguals as the only attracting population. The stable distributions in their model imply the death of one tongue. Minett and Wang suggest what strategies might be applied to sustain a language, but this discussion remains highly abstract. Other variations with bilingualism (many of them inspired by this last model) incorporate geography and social networks, as discussed above \cite{VazquezSan2010, CastelloSan2006, CastelloSan2007a, CastelloSan2007b, StaufferSan2007}. A great review of the role of social networks oriented to linguists can be found in \cite{CastelloSan2013}. Two important insights are summarized there: First, in some cases the introduction of bilingual agents tends to accelerate the extinction of one of the languages. This happens in so-called {\em small-world} networks, in which the social structure ensures that all speakers are relatively close to each other -- i.e. options or opinions spread easily across the network. This seems at odds with the results just discussed \cite{MiraParedes2005, MiraNieto2011, OteroMira2013}, in which bilingualism favors language coexistence. These differences will be solved in the next section. The second important message is that, when the network presents a strong community structure (with social clusters relatively isolated from each other but well connected within themselves), coexistence of both languages is possible if monolinguals are confined into homogeneous clusters and bilinguals occupy the interface between the communities. This, again, implies a kind of survival by segregation as the one discussed in section II.B due to geography.

    \subsection{The delicate quantification of bilingualism}
      \label{sec:2.4}

      Heinsalu, Patriarca and L\'eonard \cite{HeinsaluLeonard2014} consider the role of the attracting population. When nothing exists but $X$- and $Y$-speakers, only they can encourage you to switch groups. When bilingual speakers enter the game, they might influence your decision too. In the previous section we saw two extremes: For Mira and colleagues \cite{MiraParedes2005, MiraNieto2011, OteroMira2013}, bilinguals are as influential as monolinguals. For Minett and Wang \cite{MinettWang2008} only monolinguals affect you despite the existence of a bilingual group. Based on this latter model, Heinsalu et al. \cite{HeinsaluLeonard2014} consider attracting populations $x + \alpha b$ and $y + \beta b$ so that transition probabilities $Y \rightarrow B$ and $X \rightarrow B$ take the form $P_{YB} \sim (x + \alpha b)^a$ and $P_{XB} \sim (y + \beta b)^a$, with $\alpha, \beta \in [0, 1]$. Only monolinguals can attract speakers away from the bilingual group.

      The Minett-Wang \cite{MinettWang2008} model did not contain stable bilingual configurations. This is also the model adopted by \cite{CastelloSan2013} and \cite{CastelloSan2007b} to simulate agents in social networks, in which the presence of bilinguals accelerated the extinction of a language when the network was well connected -- hence hindering language coexistence. With the modifications by \cite{HeinsaluLeonard2014}, depending on the relative prestiges of both languages and other parameters that control the flow of speakers, stable bilingual configurations arise if $\alpha > \alpha^*$ and $\beta > \beta^*$. This indicates that, for both languages to coexist, their speakers must perceive bilinguals as representatives of the opposite tongue. This perception has to be above some threshold ($\alpha > \alpha^*$, $\beta > \beta^*$), which in turn depends on the prestiges. Another requirement of the stable bilingual distributions is that $B \rightarrow \{X, Y\}$ transitions must be rare ($B$ eventually agglutinates all speakers in this model if it is stable). This is interpreted in \cite{HeinsaluLeonard2014} as a requisite for bilinguals to have their own intrinsic advantages. Note, finally, that $X$- and $Y$-monolinguals might perceive the bilingual community differently because $\alpha$ and $\beta$ might differ. This is interesting to model real data, e.g., if an asymmetrical situation forces speakers of a tongue (but not of the other) to become bilinguals.

      Unfortunately, \cite{HeinsaluLeonard2014} is the only existing study about attracting populations. That research was performed for $a=1$, which already presents all features of the Minett-Wang model. We tested informally how introducing similar attracting populations ($x+\alpha b$ and $y+\beta b$) in Mira et al.'s model affects its dynamics. The numerical, preliminary results confirm Heinsalu et al.'s observations: decreasing $\alpha$, $\beta$ (i.e. when bilinguals are not perceived as representatives of the alternative option) hardens the conditions for stable coexistence. 

      The form of the attracting population is an important issue given the controversy around bilingualism. This affects both theoretical (``how must bilinguals enter the equations?'') and empirical studies. Most surveys rely on speakers self-assessing their use of each tongue. Methodological discrepancies exist even for studies of a same situation carried out by one same institution. Future research must clarify how the definition of bilingualism in real data affects the explanatory power of the theoretical models.

  \section{A case study: the Galician-Spanish coexistence dynamics}
    \label{sec:3}

    Galician is a romance language evolved in the northwest of the Iberian peninsula after the collapse of the Roman Empire. Galician-Portuguese first constituted a unified tongue that split in two after the political division of the Portuguese and Galician kingdoms in the XVI century. Galicia was integrated in what nowadays is Spain, whose ruling elite spoke Castillian (Spanish). Towards the end of the XIX century the contact between Galician and Spanish intensified reaching most demographic substrates, notably in the cities. This is reflected in a lasting decline \cite{RAG2004, RAG2008} starting at the turn of the XX century. This suggests that both languages were effectively segregated until the $1900$s. Similar processes happened to Basque, Catalan, and other Spanish tongues. 

    Since the early $1990$s the Real Academia Galega (Galician Royal Academy) periodically commissions surveys to assess the condition of the Galician language resulting in exhaustive sociodemographic reports that link Galician or Spanish usage to geography, age, or economic substrate. Although the same institution developed all these works, methodological differences exist between successive versions, precisely because human language is such a nontrivial feature and intense procedural and theoretical debate still exists. We focus on the age segregation of Galician and Spanish usage from the 2008 {\em Mapa socioling\"u\'istico de Galicia} (Galician sociolinguistic map). Informants were asked to classify themselves as: i) only speak Galician, ii) mostly speak Galician, iii) speak both equally, iv) mostly speak Spanish, v) only speak Spanish. We aggregated the three central groups as bilinguals to build up data for the variable $b$ discussed above, hence bilingualism is defined through {\em self-assessment of language usage} for this academic exercise. The extreme options make up $x$ and $y$ respectively. We will compare these variables to Mira et al.'s model \cite{MiraParedes2005, MiraNieto2011, OteroMira2013}. 

    These data (figure \ref{fig:4}{\bf d}) give us noisy snapshots of the real dynamical process of language shift going on. We can take the model equations \cite{MiraParedes2005}, try different values for their parameters ($a$, $c$, $k$, $s$), and select those that are in good agreement with the data -- i.e. $(a, c, k, s)$-combinations whose dynamics follow closer the empirical fractions of speakers over time. Data from the Mapa Socioling\"u\'istico de Galicia \cite{RAG2004, RAG2008} separates speakers by geographic subdivisions, so we could repeat this process for several such sub-units and also for the aggregated data. Plotting the parameters found for each region we can assess where is language coexistence {\em not} possible (red crosses in figure \ref{fig:4}{\bf c} as indicated above). Figure \ref{fig:4}{\bf d} shows extrapolations for the aggregated Galician data for which bilingualism is predicted to be a stable trait. Both monolingual groups are predicted to survive, with the Spanish monolingual group likely dominating -- however, the data is consistent with a great variability concerning the final fractions of monolingual speakers. Note the importance of having {\em quantitative} and {\em objectively reproducible} measurements for similarity and prestige.

    As we announced before, when the model in \cite{MiraParedes2005} is matched to data, $k$ can be interpreted as a similarity between the languages involved. More specifically, $k$ reflects the {\em interlinguistic similarity} as perceived by the speakers of both languages in the precise situation under research. Figure \ref{fig:4}{\bf c} shows how Galician and Spanish are perceived with an almost constant similarity across different areas ($k \sim 0.78$ with a standard deviation of $0.04$). This highlights the robustness of $k$ as an empirical measure of distance despite the great variation of the prestige parameter, which reflects very different sociolinguistic scenarios.

    Other models have been tested against real data, including the original AS-model \cite{AbramsStrogatz2003}. Language shift in the British Isles is an empirically well documented case with rich data (see \cite{KandlerSteele2010} and references therein, where a model with local birth, death, and migration is also successfully tested). Zhang and Gong's model \cite{ZhangGong2013}, as remarked above, also reproduced data with minimal empirical input. All these are predictions made by different models with the data available. This is not an indisputable truth, but rather a guideline that must be subjected to revision and that allows us to falsify the equations. It was shown \cite{HeinsaluLeonard2014} that these models are sensible to how we account for bilingualism. Besides, empirical data are updated regularly, also due to methodological advances. Statements about real-life situations must be taken with extreme care and be continuously refined. Extreme social and economic events shall invalidate our current conclusions, simply because they violate the hypotheses of the models. Locating such game-changers in real data (compared to usual statistical fluctuations) can give us a hint of the magnitude that a policy must have to support minority languages.

  \section{Discussion and prospects}
    \label{sec:4}

    This review began with theoretical considerations beyond language contact which dealt with human interactions in general: Do laws emerge for social dynamics as in physics? In what circumstances are they valid? Are there, indeed, fundamental laws? The contact scenarios reviewed are fueled by interactions between people: observing other speakers might compel us to change our tongue. Hence, what is the value of human interaction? Can we quantify it in general as we quantify the mass and charge of an electron? Powerful mathematical and empirical reasons suggest positive answers for some of these questions. This position is compatible with delicate matters about free will: intrinsic laws that guide individual actions do not exist. Instead, unavoidable regularities emerge (even for the most random systems, as thermodynamics shows) whose average trends and shifting rates become clear-cut as we aggregate larger numbers of people.

    We focus here on practical applications to language contact scenarios. This is a testing bench for the logic just outlined, and a relevant one with deep social and economic implications. The last decade has seen interesting developments, mainly hindered by the multiple factors involved and the difficult interpretation of real data \cite{Crystal2000, Wichmann2008}. Recent advances single out important factors to assess the fate of minority languages, which we summarize:

      \begin{itemize}

        \item {\bf Extinction or segregated survival as limiting cases}: This is supported by the original AS model \cite{AbramsStrogatz2003} and also by Shelling's \cite{Schelling1971} and Axelrod's \cite{Axelrod1997} account of ethnic and cultural dynamics. When interaction (hence competition) is intense, the tendency is to homogenize the cultural space. Only through weak interactions can different cultures survive, but weak interactions mean isolation. 

        \item {\bf The geographic distribution of speakers matters}: As shown in \cite{PatriarcaHeinsalu2009}, a dense stronghold of speakers of a minority language might reverse an adverse situation. But this still implies the extinction of one of the tongues. Both languages can survive if human mobility is reduced by physical barriers -- i.e., again, if languages are effectively segregated. 

        \item {\bf Bilingualism is an asset}: A naive view might suggest that a language loses speakers through the bilingual group. But some theoretical models indicate that bilinguals (when successful and salient) soften the competition and enable possibilities that were forbidden in simpler models \cite{MiraParedes2005, MiraNieto2011, OteroMira2013}. Put otherwise: pairs of languages for which a stable bilingual group cannot be successfully established are deemed to either disappear or live isolated. 

        \item {\bf Bilingualism demands two conditions to enable a steady coexistence \cite{HeinsaluLeonard2014}}: i) For the speaker of a language, bilinguals must be representative of the opposite option -- i.e. bilinguals must contribute to the attracting population. ii) The transitions away from bilingualism must be rarer than the transitions towards bilingualism, suggesting that being bilingual has attached some extra social value. This is not far fetched: bilinguals not only can engage with more people, but also bring together large, linguistically isolated groups.

      \end{itemize}

    A note of caution is in order. While testing the models gave positive results, agreement between equations and data does not mean that all the theoretical conclusions must hold always. These models are valid when some strict conditions are met, but social systems are prone to unpredictable events impossible to account for. Localizing drastic shifts in the recorded data (e.g. due to wars or migrations) can be an interesting exercise. Such events should notably alter the smooth evolutions reproduced by the equations. These could give us a quantitative reference about the magnitude that a policy must have to positively affect the fate of minority languages. More theoretical developments are necessary to incorporate such rare events in the theory while more empirical tests are necessary to determine the limits of the current models.

    New threats and opportunities arise for minority languages as the socio-economic stage changes, prominently thanks to the internet and the levels of globalization attained in the last decades. There is a chance now to test models of microscopic situations against real data gathered online. Also, modeling multilingual scenarios, studying their stability, and comparing them with real trends seems within reach.

  \bibliographystyle{chicago}

\end{document}